\documentclass[11pt, a4paper]{article}

\usepackage{mathtools}
\usepackage{amssymb}  		
\usepackage[dvipsnames, x11names]{xcolor} 

\usepackage{tikz}
\usetikzlibrary{bending, calc, decorations.markings, shapes.geometric, patterns, arrows.meta, shadings} 

\usepackage{pgfplots}
\pgfplotsset{compat=newest}

\usepgfplotslibrary{colormaps, fillbetween, external, groupplots}

\usepackage{graphicx}
\usepackage{cite}
\usepackage{authblk}

\usepackage{url}

\usepackage{hyperref}
\hypersetup{colorlinks=true, linkcolor=Blu, citecolor=Blu, urlcolor=Blu}

\tikzset{
	partial ellipse/.style args={#1:#2:#3}{insert path={+ (#1:#3) arc (#1:#2:#3)}},
}

\newcommand*{\vb}[1]{\boldsymbol{#1}}
\newcommand*{\mat}[1]{\begin{pmatrix} #1 \end{pmatrix}}

\newcommand*{\R}{\mathbb{R}}
\newcommand*{\C}{\mathbb{C}}
\newcommand*{\Sph}{\mathbb{S}}			
			
\newcommand*{\HDU}{H_{\text{D},U}}

\newcommand*{\idm}{I} 

\DeclareMathOperator{\tr}{tr}

\newcommand*{\iu}{\mathrm{i}\mkern1mu}  
\newcommand*{\e}{\mathrm{e}} 
\newcommand*{\dd}[1]{\mathrm{d}#1}

\newcommand*{\dom}{\mathfrak{D}} 
\newcommand*{\Leb}{\mathcal{L}} 
\newcommand*{\Sob}{\mathcal{H}} 
\newcommand*{\UU}{\mathrm{U}}

\definecolor{Blu}{cmyk}{1,0.6,0,0.2}	
\definecolor{Blu1}{cmyk}{0.2,0.09,0,0.1} %opacity=0.3
\definecolor{Blu2}{cmyk}{0.37,0.16,0,0.18} %opacity=0.5
\definecolor{Rosso}{cmyk}{0,1,0.8,0.2}	
\definecolor{Ocra}{cmyk}{0,0.1,1,0.2}

\begin{document}

\title{\textbf{Dimensional reduction of the Dirac equation in arbitrary spatial dimensions}}

\author[$\hspace{0cm}$]{Giuliano Angelone$^{1,2,}$\footnote{\href{mailto:giuliano.angelone@ba.infn.it}{giuliano.angelone@ba.infn.it}}}

\affil[$1$]{\small Dipartimento di Fisica, Universit\`a di Bari, I-70126 Bari, Italy}
\affil[$2$]{\small INFN, Sezione di Bari, I-70126 Bari, Italy}

%\mainmatter
	\title{\textbf{Hearing the boundary conditions of the one-dimensional Dirac operator}}

\maketitle

\begin{abstract}
We study the isospectrality problem for a relativistic free quantum particle, described by the Dirac Hamiltonian, confined in a one-dimensional ring with a junction. We analyze all the self-adjoint extensions of the Hamiltonian in terms of the boundary conditions at the junction, characterizing the energy spectrum by means of a spectral function. By determining the symmetries of the latter, we are able to divide the self-adjoint extensions in two classes, identifying all the families of isospectral Hamiltonians, and thus completely characterizing the isospectrality problem.

%\keywords{Quantum boundary conditions, self-adjoint extensions, Dirac operator, isospectrality}
\end{abstract}

\section{Introduction}
As it is well known, in any bounded region $\Omega\subset \R^2$ the Laplacian with Dirichlet BCs has a discrete spectrum, say $\{\lambda_n\}_{n\in \mathbb{N}}$, which can in principle \emph{always} be determined by solving the spectral problem
\begin{equation}
-\Delta u_n(x,y)=\lambda_n u_n(x,y)\,,\qquad u_n(x,y)|_{\partial\Omega}=0\,,
\end{equation}
assuming that $\Omega$ and its boundary $\partial\Omega$ are sufficiently regular. 
This setting emerges as an idealized model of a drum, described as a two-dimensional vibrating membrane $\Omega$ whose boundary $\partial\Omega$ is kept fixed by a frame, so that the $\lambda_n$, the eigenvalues of the Laplacian, represent its normal frequencies. In  a famous paper of 1966, Mark Kac popularized the related \emph{inverse} spectral problem, which can be formulated as follows \cite{Kac66}: can we \emph{uniquely} determine the region $\Omega$ just by knowing the full set of the corresponding  Laplace eigenvalues $\lambda_n$? Or, conversely, can we construct two \emph{isospectral} and non-isometric regions $\Omega_1$ and $\Omega_2$, i.e.~having different shapes but the same spectrum?

\begin{figure}[tb]
\centering
%\figname{isoboundary_isosp}
\begin{tikzpicture}[scale=2]
\draw[very thick, fill=Blu1] (3,0)--(3,2)--(3.5,2.5)--(4,2)--(4,2)--(3.5,1.5)--(4,1)--(4,0)--(3.5,0.5)--cycle;
\draw[dotted] (3.5,0.5)--(3,1) --(4,1) -- (3.5,0.5) 
(3,1) -- (3.5,1.5)-- (3,2)--(4,2);

\draw[very thick,  fill=Blu1] (0,0)--(0.5,0.5)--(0,1)--(0,2)--(1,2)--(0.5,1.5)--(1.5,0.5)--(1,0)--cycle;
\draw[dotted] (0.5,0.5) -- (1,0) -- (1,1)--(0.5,0.5) (1,1)--(0,1)--(0.5,1.5)--(0,2);

\node at (-0.5,1) {$\Omega_1$};
\node at (4.5,1) {$\Omega_2$};
\end{tikzpicture}
\caption{A pair of isospectral polygons, constructed with seven half-square tiles differently arranged (the dotted lines are a guide for the eye), see \cite{GoWeWo92}.}
\label{fig:isosp}
\end{figure}
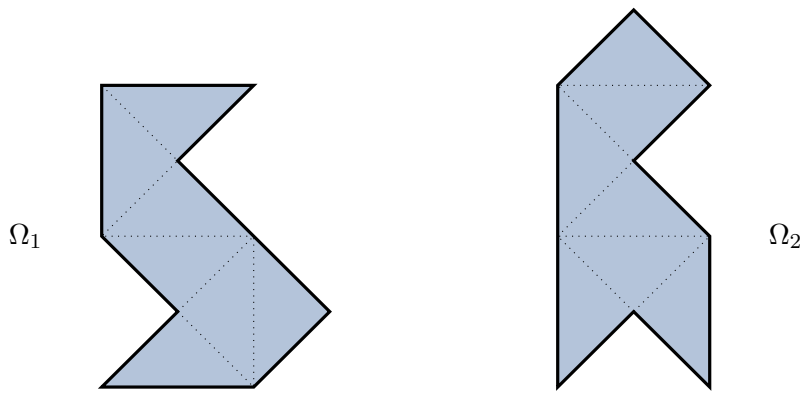

Remarkably, even after more than fifty years, this problem has been only partially solved~\cite{GiTh10, Zel04}. 
A first negative answer has been given in 1992 by C.~Gordon \emph{et al.}~\cite{GoWeWo92}, who constructed an isospectral pair of two-dimensional polygons, depicted in \autoref{fig:isosp}. 
In the case of two-dimensional domains with a smooth boundary, a positive result can be recovered by requiring some additional symmetries, see e.g.~the work of S.~Zelditch~\cite{Zel00}, but the general problem is still unsolved. 
Besides, many related questions (which can generally fall under the umbrella of ``isospectrality'', or of inverse problems) are still unsolved, and object of research. 
Here we briefly mention some interesting topics, involving the isospectrality problem for regions with a fractal boundary~\cite{BrCa86},  relativistic billiards~\cite{DiHu20}, quantum graphs~\cite{GuSm01, Rue15, LaKu20, LaKu21, FaLlPo23}, photonic systems~\cite{photo21}, and also the experimental construction of some isospectral regions~\cite{SrKu94, EvPi99}.

In this paper, we are interested in the following variation of the Kac's problem, formulated for a “quantum drum”, that is \emph{quantum billiard} consisting of a free quantum particle moving in a region of space $\Omega$: which boundary conditions, for a given domain $\Omega$, are associated with the same energy eigenvalues? In other words we fix the shape of the drum and we ask whether it is possible to distinguish different quantum boundary conditions just by looking at the energy spectrum of the particle. In \cite{isoboundary}, in particular, the authors studied the case of a non-relativistic particle (whose kinetic-energy is described by the free Schrödinger operator, i.e.~the Laplacian) moving in a one-dimensional ring with a junction. This one-dimensional system could be implemented in a SQUID, a superconducting ring where different BCs can be obtained by a flux-loop tunable junction~\cite{Vion,Poletto,Paauw}. Here, we extend the analysis of \cite{isoboundary} by considering a spin-$\tfrac{1}{2}$  \emph{relativistic} particle, described by the free Dirac operator, moving on the same domain. Note that from a physical perspective this inverse spectral problem might have applications in quantum metrology with SQUIDs~\cite{Friedman}, atoms in cavities~\cite{Haroche}, and ions and atoms in magnetic traps~\cite{Paul}. 

The paper is organized as follows. In Section~\ref{sec:nonrel} we review the isospectrality problem for a non-relativistic particle, which can be analyzed by determining a certain spectral function. In Section~\ref{sec:rel} we then extend the analysis to a relativistic particle, drawing our conclusions in Section~\ref{sec:conclusions}.

\section{Hearing the boundary conditions of a quantum billiard}\label{sec:nonrel}
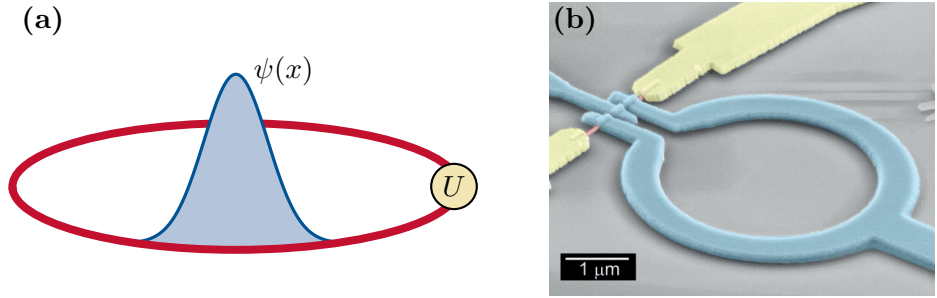
\begin{figure}[tp]
\centering
%\figname{intro_ring}
\begin{tikzpicture} %[xscale=0.9, yscale=1.1]
\pgfmathdeclarefunction{gauss}{2}{%
  \pgfmathparse{1/(#2*sqrt(2*pi))*exp(-((x-#1)^2)/(2*#2^2))}%
}

\begin{scope}[scale=0.7]
\draw[line width=3pt, Rosso] (-3.8,0) [partial ellipse=0:185:4.2cm and 1.2cm];
%\draw[very thick, samples=100, smooth, domain=-1.95:1.93, Blu, fill=Celeste, fill opacity=0.7]  plot (\x-3.8, {0.09+4*exp(-((\x)^2)/(2*0.6^2))*0.8-1.165} ) -- (-3.2,-1.2) -- (-4.8,-1.18) -- cycle;
\draw[very thick, samples=100, smooth, domain=-1.95:1.93, Blu, fill=Blu1]  plot (\x-3.8, {0.09+4*exp(-((\x)^2)/(2*0.6^2))*0.8-1.165} ) -- (-3.2,-1.2) -- (-4.8,-1.18) -- cycle;
\draw[line width=3pt, Rosso] (-3.8,0) [partial ellipse=180:365:4.2cm and 1.2cm];

\node[thick, circle, draw=black, fill=Ocra!30, inner sep=2pt] at (0.3,0) {$U$};
\node at (-2.9,2.2) {$\psi(x)$};
\end{scope}
\node at (4,0.5) {\includegraphics[width=0.4\textwidth]{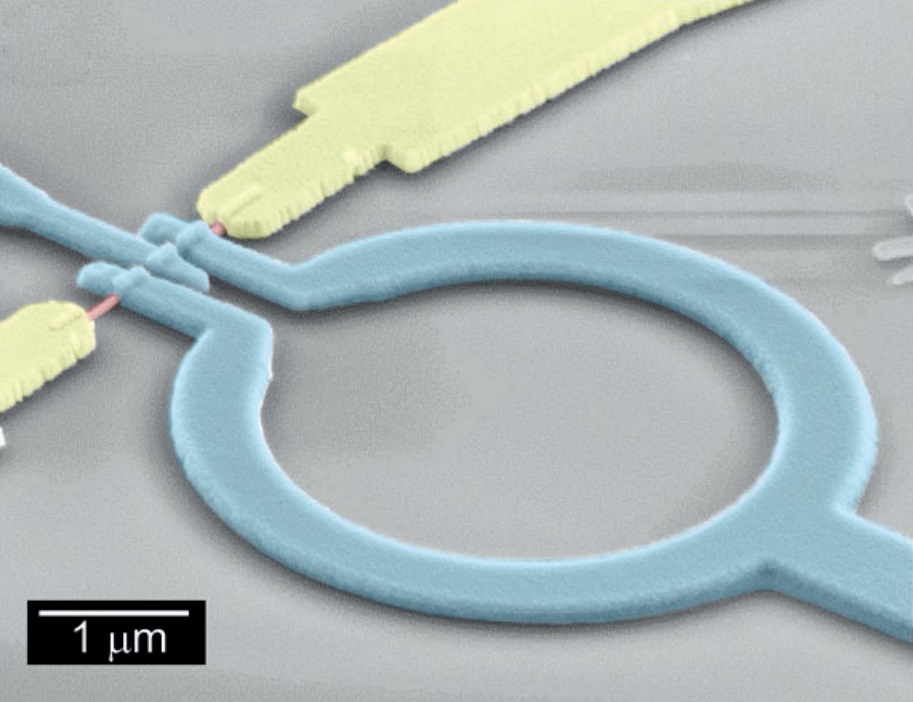}};

\node at (-5.2, 2.2) {\textbf{(a)}};
\node at (1.8, 2.2) {\textbf{(b)}};

\end{tikzpicture}

\caption{A quantum particle in a one-dimensional ring with a junction: (a) schematic diagram, where $U$ parametrizes the boundary conditions at the junction, according to Eq.~\eqref{eq:be}; (b) scanning electron microscope image of a SQUID \cite{JoMaKo17}, which represents a possible experimental realization of this system.}
\label{fig:ring}
\end{figure}

Let us consider a free non-relativistic quantum particle of mass $m$ in a ring  with a junction,  as the one depicted in \autoref{fig:ring}, idealizing the junction as a (generic) point interaction. This system, which is essentially one-dimensional, is described by the kinetic-energy operator 
\begin{equation}\label{eq:HU}
H=-\frac{\hbar^2}{2m}\frac{\mathrm{d}^2}{\mathrm{d}x^2}
\end{equation}
that is the free Schrödinger operator, acting on the Hilbert space of square integrable functions $\Leb^2(-L/2, L/2)$, where $L>0$ denotes the length of the ring. Equation~(\ref{eq:HU}) describes the action of $H$ in the ring \emph{away} from the junction. In order to generate a well-defined dynamics, the Hamiltonian $H$ should be equipped with suitable boundary conditions (BCs), which specify the behavior of the particle at the junction. In quantum mechanics the possible behaviors at the boundary, encoded in the domain $\dom(H)$ of $H$, cannot be arbitrary, but are constrained by a basic principle: $H$ must be \emph{self-adjoint}, i.e.\ $\dom (H)=\dom (H^\dagger)$ and $H=H^\dagger$. Indeed, self-adjointness is a necessary and sufficient condition for the operator to have a purely real spectrum and to generate a unitary dynamics. 

Different domains correspond to different behaviors of the particle at the junction and thus give rise to different dynamics. All the self-adjoint realizations  of the Hamiltonian \eqref{eq:HU} are known to be in one-to-one correspondence with the set of $2\times 2$ unitary matrices $U\in\UU(2)$~\cite{AIM05, AIM15}. Each of these realizations, which we henceforth denote by $H_U$, is defined on the domain
\begin{equation}
\dom(H_U)=\bigl\{\psi\in \Sob^2\bigl( -\tfrac{L}{2},\tfrac{L}{2}\bigr):\Psi_-=U\Psi_+\bigr\}\,,
\end{equation}
where $\mathcal{H}^2(\Omega)$ denotes the space of wavefunctions with square-integrable first and second derivative (i.e.~the second Sobolev space) having support on $\Omega$  and 
\begin{equation} 
\Psi_\pm= \mat{-L_0 \psi'\bigl(-\tfrac{L}{2}\bigr) \pm \iu \psi\bigl(-\tfrac{L}{2}\bigr) \\[6pt] +L_0\psi'\bigl(+\tfrac{L}{2}\bigr) \pm \iu \psi\bigl(+\tfrac{L}{2}\bigr) }
\end{equation}
are the relevant boundary values of $\psi(x)$, with $L_0$ an arbitrary scale length which we henceforth take as $L_0=L$ for convenience. In other words, to each unitary matrix $U$ there corresponds the quantum BC
\begin{equation}\label{eq:be}
\Psi_-=U\Psi_+\,,\qquad U\in\UU(2)\,,
\end{equation}
which in turn prescribes a linear relation between the vectors of boundary data $\Psi_+$ and $\Psi_-$. We mention that, at least in one-dimensional settings, there is a one-to-one mapping between self-adjoint BCs and self-adjoint (delta-like) point interactions, see e.g.~\cite{Kur96}. The expression \eqref{eq:be} admits many interesting BCs as particular cases, such as the Robin conditions
\begin{align}\label{eq:Robin}
-\psi'\bigl(-\tfrac{L}{2}\bigr)=-\tfrac{1}{L}\cot\bigl(\tfrac{\alpha}{2}\bigr)\psi\bigl(-\tfrac{L}{2}\bigr)\,, && \psi'(\tfrac{L}{2})=-\tfrac{1}{L}\cot(\tfrac{\alpha}{2})\psi(\tfrac{L}{2})\,,
\end{align}
and the pseudo-periodic conditions
\begin{align}
\psi\bigl(\tfrac{L}{2}\bigr)=\e^{\iu\alpha} \psi\bigl(-\tfrac{L}{2}\bigr)\,,&& \psi'\bigl(\tfrac{L}{2}\bigr)=\e^{\iu\alpha} \psi'\bigl(-\tfrac{L}{2}\bigr)\,,
\end{align}
obtained respectively with the unitary matrices
\begin{align}
U_\text{R}(\alpha)=\mat{\e^{\iu\alpha}&0 \\ 0&\e^{\iu\alpha}}&&\text{and}&& U_\text{pp}(\alpha)=\mat{0&-\e^{-\iu\alpha} \\ -\e^{\iu\alpha}&0}
\end{align}
for each $\alpha\in[0,2\pi)$. Before proceeding, let us observe that although the “shape” of the system is fixed,  its topology is not, being determined by the  BCs:  while non-local BCs effectively describe a particle in a ring, local conditions (associated with a diagonal matrix $U$) model instead a particle confined in a interval, i.e.~a one-dimensional box.

\subsection{Isospectrality}
The isospectrality problem for this one-dimensional system can be tackled by computing the \emph{spectral function} $F_U(E)$, whose real zeroes, for each choice of $U$, characterize the spectrum $\sigma(H_U)$ of  $H_U$:
\begin{equation}
\sigma(H_U)=\{E\in\R:F_U(E)=0 \}\,.
\end{equation}
As it turns out, the spectral function of this system takes the form
\begin{align}
F_U(E)=\det(U)-a(E)\tr(U)+b(E)\tr(U\sigma_x)+c(E)\label{eq:FU1} 
\end{align}
where $\sigma_x$ is the first Pauli matrix while $a(E)$, $b(E)$ and $c(E)$ are certain functions of the energy $E$ obtained from the solution of the spectral problem, see \cite{isoboundary} for details. In particular, the above spectral function depends on $U$ only trough the quantities $\det(U)$, $\tr(U)$ and $\tr(U\sigma_x)$, which cannot completely characterize the four-dimensional space of parameters of a generic $\UU(2)$ matrix $U$. This observation is reflected in a (boundary) symmetry of the spectral function: for any 
$\lambda\in\mathbb{R}$
we have that 
\begin{equation}\label{eq:FUFUlambda}
	F_U(E)=F_{U_\lambda}(E),
\end{equation}
 where
\begin{equation}\label{eq:Udelta}
U_\lambda= \e^{\iu\lambda\sigma_x}U\e^{-\iu\lambda\sigma_x}\,.
\end{equation}
One can easily verify that the $\UU(2)$ transformation $U\mapsto U_\lambda$ is indeed the most general one which simultaneously preserves the values of $\det(U)$, $\tr(U)$ and $\tr(U\sigma_x)$ appearing in Eq.~\eqref{eq:FU1}. 

Remarkably, the boundary symmetry \eqref{eq:FUFUlambda} is ultimately due to a symmetry of the \emph{bulk}, associated with the one-parameter unitary group 
\begin{equation}
\mathcal{P}_{\lambda}= \e^{\iu\lambda P}=\cos(\lambda) \idm+\iu\sin(\lambda) P\,,\qquad \lambda\in\mathbb{R}\,,
\end{equation}
with $I$ the $2\times 2$ identity matrix,  generated by the parity operator 
\begin{equation}
P\colon \psi(x)\mapsto\psi(-x)
\end{equation}
on $\Leb^2(- L/2, L/2)$.  For each $\lambda\in\mathbb{R}$ the action of $\mathcal{P}_{\lambda}$ leaves formally invariant the Hamiltonian~\eqref{eq:HU}, that is the Laplacian, but  changes its domain according to
\begin{equation}\label{eq:PHP}
\mathcal{P}_{\lambda} H_U \mathcal{P}_{\lambda}^{\dagger}=H_{U_\lambda}\,,
\end{equation}
with $U_\lambda$ given by Eq.~\eqref{eq:Udelta}. As a consequence of the above unitary equivalence, for each $\lambda\in\R$ we get  the following \emph{isospectral relation}:
\begin{equation}\label{eq:isospectrality}
\sigma(H_U)=\sigma\bigl(H_{\e^{\iu\lambda\sigma_x}U\e^{-\iu\lambda\sigma_x}}\bigr)\,.
\end{equation}
Moreover, by solving Eq.~\eqref{eq:Udelta} for $U=U_\lambda$, one can find the family of self-adjoint Hamiltonians which are actually symmetric under $\mathcal{P}_\lambda$ (or, equivalently, under $P$). Such operators belong to the $(\UU(1)\times \UU(1))/\mathbb{Z}_2$ subgroup  of $\UU(2)$ described by the matrices
\begin{equation}\label{eq:Uparity}
U(\eta,\theta)= \e^{\iu (\eta \idm+\theta \sigma_x)}=\e^{\iu\eta}\left(\begin{array}{@{}cc@{}}
	\cos(\theta) & \iu\sin(\theta)  \\ \iu\sin(\theta) & \cos(\theta)
\end{array}\right)\,,
\end{equation}
with $\eta\in[0,\pi)$ and $\theta\in[0,2\pi)$. 

This result completes the characterization of the isospectrality problem for this system. The answer is  only partially positive: we can indeed hear \emph{some} BCs. As it turns out, there is a two-parameters family of BCs, and thus of  Hamiltonians $H_U$, such that each element is uniquely associated with the corresponding spectrum $\sigma(H_U)$ and vice versa. These BCs,  described by the matrices $U(\eta,\theta)$ defined in Eq.~\eqref{eq:Uparity}, are associated with parity-symmetric self-adjoint realizations. Note that, for example, the Robin conditions \eqref{eq:Robin} belong to this family. Conversely, each BC $U$ not belonging to the family \eqref{eq:Uparity} generates a $\UU(1)$-family of isospectral BCs $\{U_\lambda\}_{\lambda\in\R}$, with $U_\lambda$ given by Eq.~\eqref{eq:Udelta}. The corresponding Hamiltonians, which are not symmetric under parity, are related by Eqs.~\eqref{eq:PHP} and~\eqref{eq:isospectrality} and thus they all ``sound''  the same. An example of this case is given by the $\UU(1)$-family of quasi-periodic conditions
\begin{align}
    \psi\bigl(\tfrac{L}{2}\bigr)=\iu\cot\bigl(\tfrac{\alpha}{2}+\tfrac{\pi}{4}\bigr)\psi\bigl(-\tfrac{L}{2}\bigr)\,,&&
    \psi'\bigl(\tfrac{L}{2}\bigr)=\iu\tan\bigl(\tfrac{\alpha}{2}+\tfrac{\pi}{4}\bigr)\psi'\bigl(-\tfrac{L}{2}\bigr)\,,
\end{align}
corresponding to the matrices
\begin{align}\label{eq:Uqp}
    U_\text{qp}(\alpha)=\mat{-\sin(\alpha)& \iu\cos(\alpha) \\ -\iu \cos(\alpha) & \sin(\alpha)}\,,&&\alpha\in[0,2\pi)\,,
\end{align}
and admitting as particular cases the pseudo-periodic BC  with phase $\e^{\iu\pi/2}$
\begin{align}
\psi\bigl(\tfrac{L}{2}\bigr)=\iu\psi\bigl(-\tfrac{L}{2}\bigr)\,,&&\psi'\bigl(\tfrac{L}{2}\bigr)=\iu\psi'\bigl(-\tfrac{L}{2}\bigr)\,,
\end{align}
for $\alpha=0$, and the mixed Neumann-Dirichlet BC
\begin{align}
\psi'\bigl(-\tfrac{L}{2}\bigr)=0\,,&&\psi\bigl(\tfrac{L}{2}\bigr)=0\,,
\end{align} 
for $\alpha=\pi/2$. Note in particular that
\begin{equation}
\e^{\iu\lambda\sigma_x}U_\text{qp}(\alpha)\e^{-\iu\lambda\sigma_x}=U_\text{qp}(\alpha-2\lambda)\,.
\end{equation}
As expected, a standard direct computation shows that the above particular cases share indeed the same spectrum, which is given by
\begin{equation}
\sigma\bigl(H_{U_\text{qp}(\alpha)}\bigr)=\Bigl\{ E_n=\frac{\hbar^2\pi^2}{2mL^2}\Bigl(n+\frac{1}{2}\Bigr)^2:n\in \mathbb{N}\Bigr\}\,.
\end{equation}
\section{A relativistic particle in a ring with a junction}\label{sec:rel}
We now consider a relativistic (spin-$\tfrac{1}{2}$) quantum particle, of mass $m$, free to move in the one-dimensional ring  with a junction considered in the previous section. In this case, the relevant kinetic-energy operator is given by the one-dimensional free Dirac operator \cite{Tha92}
\begin{equation}
H_\text{D}=-\iu\hbar c \alpha \frac{\dd }{\dd{x}}+mc^2\beta 
\end{equation}
acting on $\Leb^2(-L/2,L/2)\otimes \C^2$, where $c$ is the speed of light while $\alpha$ and $\beta$ are two $2\times2$ Hermitian matrices such that\footnote{This is the relevant operator that is obtained from the ususal three-dimensional Dirac Hamiltonian by means of dimensional reduction, see e.g.~\cite{diracred}.}
\begin{align}\label{eq:alphabeta}
\alpha\beta+\beta\alpha=0\,,&&\alpha^2=\beta^2=I\,.
\end{align}
Each particular choice of these two matrices is usually called a \emph{representation} of the Dirac operator. From the physical perspective the latter is somewhat immaterial, since any two Dirac Hamiltonians, say $H_\text{D}$ and $\tilde{H}_\text{D}$, corresponding to two different sets of matrices $\{\alpha,\beta\}$ and $\{\tilde{\alpha},\tilde\beta\}$, are related by the unitary transformation $\tilde{H}_\text{D}=VH_\text{D}V^\dagger$ where $V$ is the unitary matrix such that \cite{Good55}
\begin{align}\label{eq:V}
V\alpha V^\dagger=\tilde{\alpha}\,,&&V\beta V^\dagger=\tilde{\beta}\,.
\end{align}
In any case, for definiteness we henceforth consider the Dirac (or standard) representation $\alpha=\sigma_x$ and $\beta=\sigma_z$, that is we fix the Dirac operator to be
\begin{equation}
H_\text{D}=-\iu\hbar c \sigma_x \frac{\dd }{\dd{x}}+mc^2\sigma_z\,.
\end{equation}

As explained in the previous section, we need to supply suitable BCs in order to consider a proper self-adjoint realization of the Dirac operator, see also \cite{AloDeV97, BeSaWi02, AIM15}. In Appendix~\ref{sec:bcs} we construct all the self-adjoint extensions of $H_\text{D}$ by using a representation-independent approach based on the boundary triple machinery. As it turns out, also in the relativistic case the self-adjoint extensions of the Hamiltonian are in one-to-one correspondence with the unitary matrices $U\in\UU(2)$, and we denote them by $H_{\text{D},U}$. By further denoting the \emph{spinorial components} of a generic wavefunction as 
\begin{equation}
\Psi(x)=\bigl(\phi(x) \quad \chi(x)\bigr)^{\intercal}\,,
\end{equation} 
the domain of $H_{\text{D},U}$ can be written as
\begin{equation} %\label{eq:DHrelU}
\dom(H_{\text{D},U})=\bigl\{\Psi=\bigl(\phi \quad \chi\bigr)^{\intercal}\in \Sob^1\bigl(-\tfrac{L}{2},\tfrac{L}{2}\bigr)\otimes \C^2: \Psi_{\text{D},-}=U\Psi_{\text{D},+}\bigr\}\,,
\end{equation}
where $\Sob^1(\Omega)$ is the first Sobolev space while the vectors $\Psi_{\text{D},\pm}$ of boundary data are now given by
\begin{align}\label{eq:GpmD}
\Psi_{\text{D},\pm}= \mat{\phi\bigl(-\tfrac{L}{2}\bigr)\mp \chi\bigl(-\tfrac{L}{2}\bigr) \\[6pt] \phi\bigl(+\tfrac{L}{2}\bigr)\pm \chi\bigl(+\tfrac{L}{2}\bigr) }\,,
\end{align}
so that the BCs encoded in $\dom(H_{\text{D},U})$ explicitly read:
\begin{align}\label{eq:DirBCs}
\mat{\phi\bigl(-\tfrac{L}{2}\bigr) + \chi\bigl(-\tfrac{L}{2}\bigr) \\[6pt] \phi\bigl(+\tfrac{L}{2}\bigr) - \chi\bigl(+\tfrac{L}{2}\bigr) }=U \mat{\phi\bigl(-\tfrac{L}{2}\bigr) - \chi\bigl(-\tfrac{L}{2}\bigr) \\[6pt] \phi\bigl(+\tfrac{L}{2}\bigr) + \chi\bigl(+\tfrac{L}{2}\bigr) }\,,&&U\in\UU(2)\,.
\end{align}
To make some examples, the diagonal matrix 
\begin{align}
U_\text{ch}(\alpha)= \mat{\e^{\iu\alpha}&0 \\ 0&\e^{\iu\alpha}}\,,&&\alpha\in[0,2\pi)\,,
\end{align}
leads now to the local \emph{chiral} BCs \cite{JaMa89, HoTo96,RoYa21}
\begin{align}
\chi\bigl(-\tfrac{L}{2}\bigr)=\iu\tan\bigl(\tfrac{\alpha}{2}\bigr)\phi\bigl(-\tfrac{L}{2}\bigr)\,,&&\chi\bigl(\tfrac{L}{2}\bigr)=-\iu\tan\bigl(\tfrac{\alpha}{2}\bigr)\phi\bigl(\tfrac{L}{2}\bigr)\,.
\end{align}
These conditions relate, on the boundary, states with negative chirality with those with positive chirality, the latter being associated with wavefunctions that are respectively eigenvectors of $\sigma_x$ with eigenvalue $-1$ and $+1$.\footnote{The Pauli matrix $\sigma_x$ plays here the role of the chiral gamma matrix $\gamma^5$ \cite{JaMa89, diracred}.}  The anti-diagonal matrix
\begin{align}
\tilde{U}_\text{pp}=\mat{0&\e^{-\iu\alpha} \\ \e^{\iu\alpha}&0}\,,&&\alpha\in[0,2\pi)\,,
\end{align}
give instead the non-local BCs
\begin{align}
\phi\bigl(\tfrac{L}{2}\bigr)=\e^{\iu\alpha} \phi\bigl(-\tfrac{L}{2}\bigr)\,,&& \ \chi\bigl(\tfrac{L}{2}\bigr)=\e^{\iu\alpha} \chi\bigl(-\tfrac{L}{2}\bigr)\,.
\end{align}

In the following, after deriving in Subsection~\ref{sec:specprob} the spectral function of $\HDU$ by explicitly solving its eigenvalue equation, in  Subsection~\ref{sec:sfD} we discuss the isospectrality problem by determining the symmetries of the spectral function.

\subsection{Spectral problem} \label{sec:specprob}
The eigenvalue equation $(H_{\text{D},U}-E)\Psi(x)=0$ consists of two coupled differential equations, given by
\begin{align}\label{eq:eigenDirac}
	-\iu\phi'(x;\epsilon)=(\epsilon+\epsilon_0) \chi(x;\epsilon)\,,&&
 	-\iu\chi'(x;\epsilon)=(\epsilon-\epsilon_0) \phi(x;\epsilon)\,,
\end{align}
where here  $\epsilon=E/(\hbar c)$ and $\epsilon_0=mc/\hbar$ denote the rescaled energy and rest (or mass) energy, respectively. The space of solutions for a given $\epsilon\in\R$ is two-dimensional, and the general solution  can be written as
\begin{equation}\label{eq:gensolD}
\Psi(x; \epsilon)=c_1\mat{\phi_1(x; \epsilon)\\[6pt] \chi_1(x; \epsilon)}+c_2\mat{\phi_2(x; \epsilon)\\[6pt] \chi_2(x; \epsilon)}\,,
\end{equation}
where $c_1,c_2\in\C$. In order to solve  Eq.~\eqref{eq:eigenDirac} with the proper BCs, we note that by inserting the above solution in the expression \eqref{eq:GpmD} of the boundary values $\Psi_{\text{D},\pm}$ and by setting
\begin{equation}
\Psi_{\text{D},\pm}=A_\pm(\epsilon)\mat{
c_1 \\ c_2
}\,,
\end{equation}
where
\begin{equation}\label{eq:ApmD}
A_\pm(\epsilon)= \mat{ 
\phi_1\bigl(-\frac{L}{2};\epsilon\bigr) \mp  \chi_1\bigl(-\frac{L}{2};\epsilon\bigr)
 & \phi_2\bigl(-\frac{L}{2};\epsilon\bigr) \mp  \chi_2\bigl(-\frac{L}{2};\epsilon\bigr) \\[8pt]
\phi_1\bigl(+\frac{L}{2};\epsilon\bigr) \pm  \chi_1\bigl(+\frac{L}{2};\epsilon\bigr)
 &  \phi_2\bigl(+\frac{L}{2};\epsilon\bigr) \pm  \chi_2\bigl(+\frac{L}{2};\epsilon\bigr)}\,,
\end{equation}
imposing the BCs \eqref{eq:DirBCs} corresponds to requiring the  vanishing of the following spectral function:
%Thus we obtain that, for non-trivial $\Gamma_\pm\psi$,
\begin{align} %\label{eq:PUP}
F_{\text{D},U}(\epsilon)= \det\bigl(B(\epsilon)-U\bigr)=0\,, &&
B(\epsilon)= A_-^{}(\epsilon)A_+^{-1}(\epsilon)\,.
\end{align}
Thus, also the spectrum of the Dirac Hamiltonian can be obtained by finding the real zeros of the corresponding spectral function:
\begin{equation}
\sigma(H_{\text{D},U})=\biggl\{E\in\R : F_{\text{D},U}\biggl(\frac{E}{\hbar c}\biggr)=0\biggl\} \,.
\end{equation}
Before finding the explicit solution of the spectral problem, let us note that we can manipulate the expression of $F_{U,\text{D}}(\epsilon)$ by making use of the relation 
\begin{equation}
\det(M-N)=\det(M)+\det(N) + \tr(MN) - \tr(M)\tr(N)\,,
\end{equation}
which is generally true for any pair of $2\times 2$ matrices~\cite{BFV01}, obtaining that
\begin{equation}\label{eq:FU}
F_{\text{D},U}(\epsilon)=\det(B(\epsilon))+\det(U)+\tr(B(\epsilon)U)-\tr(B(\epsilon))\tr(U)\,.
\end{equation}

The structure of the eigenvalue equation \eqref{eq:eigenDirac}, and hence that of its solution~\eqref{eq:gensolD}, depends on the value of the wavenumber
\begin{equation}
k=k(\epsilon)= \e^{\iu \arg(\epsilon^2_{}-\epsilon_0^2)/2} \sqrt{\lvert\epsilon^2_{}-\epsilon_0^2\rvert}
\end{equation}
which is a real number outside of the mass gap, that is for $|\epsilon|\ge \epsilon_0$, and purely imaginary inside of the mass gap, i.e. for $\epsilon\in(-\epsilon_0,\epsilon_0)$. In particular, one should distinguish the analysis of the \emph{mass modes}, that is the solutions with $\epsilon=\pm \epsilon_0$ (so that $k=0$), from the other modes having a non-vanishing wavenumber. When $\epsilon=\epsilon_0$ or $\epsilon=-\epsilon_0$ we get, respectively, the coupled equations
\begin{align}
\begin{cases}
	-\iu \phi'(x;\epsilon_0)=2\epsilon_0 \chi(x;\epsilon_0)\\
 	-\iu\chi'(x;\epsilon_0)=0
\end{cases}\,,&& 
\begin{cases}
	-\iu\phi'(x;-\epsilon_0)=0\\
 	-\iu\chi'(x;-\epsilon_0)=-2\epsilon_0 \phi(x;-\epsilon_0)
\end{cases}\,.
\end{align}
The above systems have, respectively, the following solutions 
\begin{align}
\Psi(x; \epsilon_{0})=\mat{c_1+c_2 x/L\\[4pt] -\iu c_2/(2\epsilon_{0}L)  }\,,&& \Psi(x; -\epsilon_{0})=
\mat{\iu c_2/(2\epsilon_0 L)  \\[4pt]  c_1+c_2 x/L}\,,
\end{align}
from which we obtain the matrices
\begin{align}
A_{\pm}(\epsilon_0)=\mat{1 & -\frac{1}{2}\bigl(1\mp\frac{\iu}{\epsilon_0 L}\bigr) \\[6pt] 1 & +\frac{1}{2}\bigl(1\mp\frac{\iu}{\epsilon_0 L}\bigr)}\,,
&&
A_{\pm}(-\epsilon_0)=\mat{\mp 1 & \frac{1}{2}\bigl(\frac{\iu}{\epsilon_0 L}\pm 1\bigr) \\[6pt] \pm1 & \frac{1}{2}\bigl(\frac{\iu}{\epsilon_0 L}\pm 1\bigr)}\,.
\end{align}
Note that each of the above matrices is invertible, thus we compute
\begin{equation}
B(\pm \epsilon_0)=\pm\frac{1}{\epsilon_{0} L\mp \iu }\mat{\epsilon_0 L & - \iu \\[4pt] -\iu & \epsilon_0 L}\,.
\end{equation}

When $\epsilon\neq \pm\epsilon_0$, instead, we can rearrange the equations \eqref{eq:eigenDirac} as  
\begin{align}\label{eq:eigenDirac2}
\phi''(x;\epsilon)=-k^2 \phi(x;\epsilon)\,,&&
 \chi(x;\epsilon)=\frac{-\iu}{\epsilon+\epsilon_0}\phi'(x;\epsilon) \,,
\end{align}
from which we readily obtain the general solution
\begin{equation}
\Psi(x; \epsilon)=\mat{ c_1 \e^{\iu k x}+c_2 \e^{-\iu k x}\\[6pt]  \sqrt{\frac{\epsilon-\epsilon_0}{\epsilon+\epsilon_0}} (c_1 \e^{\iu k x}-c_2 \e^{-\iu k x})}\,,
\end{equation}
and the corresponding matrices
\begin{equation}
A_{\pm}(\epsilon)=
\mat{
\e^{-\iu kL/2 }\Bigl(1\mp \sqrt{\frac{\epsilon-\epsilon_0}{\epsilon+\epsilon_0}}\, \Bigr) &
\e^{+\iu kL/2 }\Bigl(1\pm \sqrt{\frac{\epsilon-\epsilon_0}{\epsilon+\epsilon_0}}\, \Bigr)  \\[6pt]
\e^{+\iu kL/2 }\Bigl(1\pm \sqrt{\frac{\epsilon-\epsilon_0}{\epsilon+\epsilon_0}}\, \Bigr)   & 
\e^{-\iu kL/2 }\Bigl(1\mp \sqrt{\frac{\epsilon-\epsilon_0}{\epsilon+\epsilon_0}}\, \Bigr)  }\,.
\end{equation}
As the reader can check by inspecting
\begin{equation}
\det\bigl(A_{\pm}(\epsilon)\bigr)=\frac{-4\iu }{\epsilon+\epsilon_0}\Bigl[\epsilon \sin\bigl(kL\bigr)\mp \iu k\cos\bigl(kL\bigr)\Bigr]\,,
\end{equation}
both $A_\pm(\epsilon)$ are invertible if and only if $\epsilon\neq  \pm\epsilon_0$. Under such assumption we can thus compute
\begin{equation}\label{eq:BED}
B(\epsilon)=a_\text{D}(\epsilon)\idm+b_\text{D}(\epsilon)\sigma_x\,,
\end{equation}
where we defined
\begin{align}
a_\text{D}(\epsilon)&= \frac{\epsilon_0 \sin(kL)}{\epsilon \sin(kL)- \iu k\cos(kL)}\,,\\
b_\text{D}(\epsilon)&=\frac{- \iu k}{\epsilon \sin(kL)- \iu k\cos(kL)}\,.
\end{align}
Similarly to the non-relativistic case \cite{isoboundary}, although the limit matrices
\begin{align}
\lim_{\epsilon\to +\epsilon_0} A_\pm(\epsilon)&=\mat{1 & 1 \\ 1 & 1}\,,\\
\lim_{\epsilon\to -\epsilon_0^{-}} A_\pm(\epsilon)&= \mat{\mp\infty & \pm\infty\ \\
\pm\infty & \mp\infty}=
\lim_{\epsilon\to -\epsilon_0^{+}} \iu A_\pm(\epsilon)\,,
\end{align}
are clearly not invertible and a fortiori different from the matrices $A_{\pm}(\epsilon_0)$ and $A_{\pm}(-\epsilon_0)$, the other two limits
\begin{equation}
\lim_{\epsilon\to \pm \epsilon_0} B(\epsilon)=B(\pm \epsilon_0)
\end{equation}
coincide with the corresponding quantities obtained for the mass modes. 

\subsection{Symmetries of the spectral function}\label{sec:sfD}
At this point, by inserting the explicit expression  \eqref{eq:BED} of $B(\epsilon)$ in the formula \eqref{eq:FU} of $F_{\text{D},U}(\epsilon)$ we obtain the final result
\begin{align}
F_{\text{D},U}(\epsilon)&=\det(U)-a_\text{D}(\epsilon)\tr(U)+b_\text{D}(\epsilon)\tr(U\sigma_x)+c_\text{D}(\epsilon) %\\
%&=\e^{2\iu\eta}-2\e^{\iu\eta}m_0a(\epsilon)+2\iu\e^{\iu\eta}m_1 b(\epsilon)+c(\epsilon)
\end{align}
where % in the second line we used Eq.~\eqref{eq:Uquant} while
\begin{align}
c_\text{D}(\epsilon)&= \det(B(\epsilon))=a_\text{D}^2(\epsilon)-b_\text{D}^2(\epsilon)
=\frac{\epsilon \sin(kL)+\iu k\cos(kL)}{\epsilon \sin(kL)- \iu k\cos(kL)}
\,.
\end{align}
As the reader can see, the isospectrality problem of the Dirac Hamiltonian mirrors that of the Schrödinger Hamiltonian: also in this case the spectral function, and hence the spectrum, depends on $U$ only through $\det(U)$, $\tr(U)$ and $\tr(U\sigma_x)$. The emerging isospectrality can be explained in terms of a symmetry of the Dirac Hamiltonian, analogously to the non-relativistic case. Here, the role of parity is played by the operator
\begin{equation}\label{eq:Parity}
\sigma_z P\colon \mat{\phi(x)\\[4pt] \chi(x)}\mapsto \mat{\phi(-x)\\[4pt] -\chi(-x)}\,,
\end{equation}
which generates the one-parameter unitary group 
\begin{equation}
\mathcal{P}_{\text{D},\lambda}= \e^{\iu\lambda \sigma_z P}=\cos(\lambda) \idm+\iu\sin(\lambda) \sigma_z P\,,\qquad \lambda\in\mathbb{R}\,.
\end{equation}
Although the latter leaves formally invariant the Dirac Hamiltonian, it  changes its domain according to
\begin{equation} \label{eq:PHP2}
\mathcal{P}_{\lambda} H_{\text{D},U} \mathcal{P}_{\lambda}^{\dagger}=H_{\text{D},U_\lambda}\,,
\end{equation}
with $U_\lambda$ given by Eq.~\eqref{eq:Udelta}, namely
\begin{equation} %\label{eq:Udelta}
U_\lambda= \e^{\iu\lambda\sigma_x}U\e^{-\iu\lambda\sigma_x}\,.
\end{equation}
As a consequence, for each $\lambda\in\R$ we get  the following isospectral relation,
\begin{equation} %\label{eq:isospectrality}
\sigma(H_{\text{D},U})=\sigma(H_{\text{D}, \e^{\iu\lambda\sigma_x}U\e^{-\iu\lambda\sigma_x}})\,,
\end{equation}
which fully characterize the isospectrality problem for the Dirac Hamiltonian. In particular, the BCs that can be heard are the same, in terms of the unitary matrices $U$ to which they correspond, that one finds in the non-relativistic case: 
\begin{equation} \label{eq:Uparity2}
U(\eta,\theta)= \e^{\iu (\eta \idm+\theta \sigma_x)}=\e^{\iu\eta}\left(\begin{array}{@{}cc@{}}
	\cos(\theta) & \iu\sin(\theta)  \\ \iu\sin(\theta) & \cos(\theta)
\end{array}\right)\,,
\end{equation}
with $\eta\in[0,\pi)$ and $\theta\in[0,2\pi)$, compare with Eq.~\eqref{eq:Uparity}.

To complete this section, we make two observations. The first one concerns the mass modes. By parametrizing $U\in\UU(2)$ as
\begin{equation}
U %=U(\eta, m_0,\vb{m})=
=\e^{\iu\eta}(m_0 \idm+\iu\vb{m}\cdot \vb{\sigma})=\e^{\iu\eta} \mat{m_0+\iu m_3 & m_2+\iu m_1  \\ -m_2+\iu m_1 & m_0-\iu m_3}\,,
\end{equation}
where $\vb{\sigma}=(\sigma_x, \sigma_y, \sigma_z)$, $\eta\in[0,\pi)$ while the four real parameters $m_0$ and $\vb{m}=(m_1,m_2,m_3)$ are subjected to the $\Sph^3$  constraint $m_0^2+|\vb{m}|^2=1$, we have that
\begin{align}\label{eq:Uquant}
\det (U)=\e^{\iu 2 \eta}\,,&& \tr (U)= 2\,\e^{\iu\eta}m_0\,,&&\text{and}&& \tr (U\sigma_x)=2\iu\e^{\iu\eta}m_1\,.
\end{align}
Accordingly, we find that the mass modes $\epsilon=\pm \epsilon_0$ belong to the spectrum $\sigma(H_{\text{D},U})$ if and only if,  respectively, 
\begin{equation}
m_1 + \sin(\eta) = \epsilon_0 L\bigl(m_0 \mp \cos(\eta) \bigr)\,.
\end{equation}
The second observation regards instead the relativistic spectral space, that is the set of \emph{distinct} spectra
\begin{equation}\label{eq:specspace}
\Sigma_\text{D}=\{\sigma(H_{\text{D},U}) \,:\, U\in\UU(2)\}\,.
\end{equation}
We mention that, exactly as it happens in the non-relativistic case, $\Sigma_\text{D}$ is given by a twisted solid torus, inheriting thus the non-trivial topology of the group $\UU(2)$, see \cite{isoboundary} for a detailed discussion. Remarkably, the non-trivial topology of $\Sigma_\text{D}$ signals the presence of spectral anholonomies and geometric phases \cite{ShaWil89, TsFuCh01}.

\section{Conclusions}\label{sec:conclusions}
In the previous section we completely characterized the isospectrality problem for the Dirac Hamiltonian on the ring with arbitrary self-adjoint BCs. In particular, we determined all the BCs which can be heard, being uniquely associated with the spectrum of the corresponding Hamiltonian. These turned out to define all the self-adjoint extensions which are symmetric under the parity operator introduced in Eq.~\eqref{eq:Parity}, and they are given by the two-parameter family \eqref{eq:Uparity2}. All the extensions which are not symmetric under parity, instead, are associated with a $\UU(1)$ family of isospectral (actually, unitarily equivalent) Dirac Hamiltonians, according to Eq.~\eqref{eq:PHP2}. Remarkably, this result exactly mirrors the one that has been found for the non-relativistic Schrödinger Hamiltonian in \cite{isoboundary}. This may be a general feature related to the fact that the Dirac Hamiltonian “reduces” to the Schrödinger Hamiltonian in the non-relativistic limit, see e.g.\cite{Tha92,AloDeV97}. Note however that analogous systems with a different domain $\Omega$ present different isospectrality structures \cite{TsFuCh01,isoboundary}.

An interesting question, which is still open, is if it is possible to discern the various energy spectra belonging to the same isospectrality family by using other physical observables (that is, other self-adjoint operators) beside the Hamiltonian. We may investigate this problem in the future, both in the non-relativistic and the relativistic setting. Another interesting extension of our analysis would be to consider higher dimensional drums $\Omega\subset\R^n$ with $n>1$. In this case, indeed, the defect indices of the Hamiltonian are typically infinite, and the unitary matrix characterizing its self-adjoint extensions is actually a unitary operator acting on a infinite-dimensional Hilbert space.

\section*{Acknowledgments}
We thank Professors Paolo Facchi and Giuseppe Marmo for useful discussions. This work was partially supported by Istituto Nazionale di Fisica Nucleare (INFN) through the project “QUANTUM”,  by Regione Puglia and QuantERA ERA-NET Cofund in Quantum Technologies (Grant No. 731473), project PACE-IN, and by the Italian National Group of Mathematical Physics (GNFM-INdAM).

\appendix
\section{Boundary conditions for the Dirac Hamiltonian}\label{sec:bcs}
In this Appendix we derive all the possible BCs which render the Dirac Hamiltonian
\begin{equation}
H_{\text{D}}=-\iu\hbar c \alpha \frac{\dd{}}{\dd{x}}+mc^2\beta
\end{equation}
a self-adjoint operator, for an arbitrary pair of $2\times2$ Hermitian matrices $\alpha$ and $\beta$ satisfying the properties \eqref{eq:alphabeta}. We will adopt the standard procedure related to the boundary triple formalism, see e.g.~\cite{BrGePa08,deO09,Sch12}, also partially following the construction of \cite{ABP13,AIM15}. When defined on $C_0^{\infty}(-L/2,L/2)\otimes \C^2$, $H_{\text{D}}$ is symmetric but not self-adjoint, $H_{\text{D}}^\dagger$ being defined on $\Sob^1(-L/2,L/2)\otimes \C^2$. An easy calculation shows that its deficiency indices $n_\pm=\dim\ker(H_{\text{D}}^\dagger\pm \iu)$ are both $n_\pm=2$. To see this, one can work in a specific representation (such as the Dirac one $\alpha=\sigma_x$ and $\beta=\sigma_z$) to compute the indices, and then invoke the unitary equivalence \eqref{eq:V} to come back to the general case. As a consequence, the self-adjoint extensions of $H_\text{D}$ can be parametrized by a $2\times 2$ unitary matrix $U\in\UU(2)$, and will be henceforth denoted by $\HDU$. Moreover, by setting
\begin{align}
\Psi_1(x)=\mat{\phi_1(x) \\[4pt] \chi_1(x)}\,,&& \Psi_2(x)=\mat{\phi_2(x) \\[4pt] \chi_2(x)}\,,
\end{align}
we can express the boundary form of $H_\text{D}$ in the following way:
\begin{align}
\Lambda(\Psi_1,\Psi_2)&=\langle H_\text{D}^\dagger \Psi_1|\Psi_2\rangle _{\mathcal{H}}-\langle \Psi_1|H_\text{D}^\dagger \Psi_2\rangle_{\mathcal{H}}\\
&=\iu\hbar c\int_{-L/2}^{+L/2} %\Psi_1^{\prime\dagger}(x)\alpha\Psi_2^{}(x)+ \Psi_1^{\dagger}(x)\alpha\Psi_2'(x)
\langle \Psi'_1(x)|\alpha\Psi_2(x) \rangle_{\C^2}+\langle \Psi_1(x)|\alpha\Psi'_2(x) \rangle_{\C^2} 
\,\dd{x}\\
&=\iu\hbar c %\bigl[
%\phi_1^{\ast}(x)\chi_2(x) +\chi_1^{\ast}(x)  \phi_2(x)\bigr]_{-L/2}^{+L/2}
\sum_{s=\pm\frac{L}{2}}\langle \Psi_1(s)|\alpha_\text{n}(s)\Psi_2(s) \rangle_{\C^2}
%\langle \Psi_1(x)|\alpha(x)\Psi_2(x) \rangle_{\C^2}|_{x=\frac{L}{2}}+\langle \Psi'_1(x)|\alpha\Psi_2(x) \rangle_{\C^2 }\bigr]
\,,
\end{align}
where $\mathcal{H}=\Leb^2(-L/2,L/2)\otimes \C^2$ and $\alpha_\text{n}(\pm L/2)=\pm\alpha$. Since $\alpha^2=I$, we have the spectral decomposition
\begin{equation}\label{eq:epem}
\alpha_\text{n}(s)=|e_+(s)\rangle \langle e_+(s)|-|e_-(s)\rangle \langle e_-(s)|
\end{equation}
where the (normalized) eigenvectors $e_\pm(s)\in\C^2$ satisfy
\begin{equation}\label{epm}
\alpha_\text{n}(s)e_\pm(s)=\pm e_\pm(s)\,.
\end{equation}
At this point we define the maps
\begin{align}\label{eq:Gpmrel}
\Gamma_\pm\colon\Sob^1\bigl(-\tfrac{L}{2},\tfrac{L}{2}\bigr)\otimes \C^2\to\C^2\,, &&
\Psi(x)\mapsto \mat{
\bigl\langle \e_\pm\bigl(-\tfrac{L}{2}\bigr)\big| \Psi\bigl(-\tfrac{L}{2}\bigr)\bigr\rangle_{\C^2}\\[6pt]
\bigl\langle \e_\pm\bigl(+\tfrac{L}{2}\bigr)\big| \Psi\bigl(+\tfrac{L}{2}\bigr)\bigr\rangle_{\C^2}
 }\,.
\end{align}
which, since $e_\pm(-L/2)=e_\mp(+L/2)= e_\mp$, simplify to
\begin{equation}\label{eq:Gpmrel2}
\Gamma_\pm\Psi(x)= \mat{
\bigl\langle \e_\mp \big| \Psi\bigl(-\tfrac{L}{2}\bigr)\bigr\rangle_{\C^2}\\[6pt]
\bigl\langle \e_\pm \big| \Psi\bigl(+\tfrac{L}{2}\bigr)\bigr\rangle_{\C^2}
 }\,.
\end{equation}
A direct calculation shows then  that
\begin{align}
\Lambda(\Psi_1,\Psi_2)/c=\langle\Gamma_-\Psi_1|\Gamma_-\Psi_2\rangle_{\C^2}-\langle \Gamma_+\Psi_1|\Gamma_+\Psi_2\rangle_{\C^2}\,,&& c=- \iu \hbar c\,,
\end{align}
i.e.~that $(\C^2,\Gamma_-,\Gamma_+)$ is a boundary triple for the Dirac Hamiltonian $H_\text{D}$, in the sense of Definition~7.1.11 of~\cite{deO09}. We can then conclude, by e.g.~Theorem~7.1.13 of~\cite{deO09}, that for each $U\in\UU(2)$ the  self-adjoint extension $\HDU$ is defined in the domain
\begin{equation}\label{eq:DHrelU}
\dom(\HDU)=\{\Psi\in \Sob^1\bigl(-\tfrac{L}{2},\tfrac{L}{2}\bigr)\otimes \C^2: \Gamma_-\Psi=U\Gamma_+\Psi\}\,.
\end{equation}
The relation $\Gamma_-\Psi=U\Gamma_+\Psi$ represents thus an $\alpha$-depending BC, expressed explicitly  in terms of the eigenvectors of $\alpha$ by Eq.~\eqref{eq:Gpmrel2}. In particular, for $\alpha=\sigma_x$ we have the eigenvectors $e_\pm=\tfrac{1}{\sqrt{2}}\mat{1&\pm 1}^\intercal$, so that the quantities
\begin{equation}
\Gamma_\pm\mat{\phi(x)\\ \chi(x)}=\frac{1}{\sqrt{2}}\mat{\phi\bigl(-\tfrac{L}{2}\bigr)\mp \chi\bigl(-\tfrac{L}{2}\bigr)\\[6pt]
\phi\bigl(+\tfrac{L}{2}\bigr)\pm \chi\bigl(+\tfrac{L}{2}\bigr)
}
\end{equation}
correspond, beside a constant multiplicative factor, to the boundary data vectors $\Psi_{\text{D},\pm}$ introduced in Eq.~\eqref{eq:GpmD}.

\end{document}